\documentstyle[12pt]{article}
\begin{document}
\title{FOUR ISSUES IN  CORRELATIONS AND FLUCTUATIONS   }
\author{A. BIALAS \\
 M.Smoluchowski Institute of Physics, Jagellonian
University \\ Reymonta 4, 30-059 Krakow, Poland \\ E-mail:
bialas@thrisc.if.uj.edu.pl} 
\maketitle
\begin{abstract}
Introductory talk at the Session on Correlations and Fluctuations of the 
XXVIIth International Symposium 
 on Multiparticle Dynamics held at Delphi, Greece from
6th till 11th of September 1998.
\end{abstract}

\section{ Introduction}

The purpose of this talk, as I understand it, is to introduce briefly
the subject of the Session to the participants who are not experts in
the field. In trying to do this I shall heavily borrow from my recent
summary of the Matrahaza workshop \cite{abm}. Since I am now much less
constrained by the program of the meeting, however, this account shall
reflect more adequately my personal views on the subject.  I
restrict myself to the four issues:

(i) Bose-Einstein interference;

(ii) Intermittency;

(iii) QCD and multiparticle correlations;

(iv) Event-by-Event fluctuations.

\section {Bose-Einstein interference}

The  discussion of  correlations 
in multiparticle production is at present largely dominated
by effects of
the Bose-Einstein interference.  
  Let me thus start by a
brief reminder what is this all about\footnote{The  physics of 
Bose-Einstein correlations  was recently extensively reviewed by G.Baym \cite{by}}.

The practical problem we face can be formulated as follows: given a
calculation (or a model) which ignores identity of particles,
the question is how to
"correct" it in order to take into account the effects of quantum
interference (which is the consequence of  identity\footnote{ It
should be understood that this problem is very common in quantum
mechanical calculations, as illustrated, e.g., by evaluation of Feynman
diagrams. I would like to thank J.Pisut and K.Zalewski for discussions
of this question. }). Let us thus suppose that we have an amplitude for
production of N particles $M^{(0)}_N(q)$ ($q=q_1,...q_N$), calculated
with the identity of particles being ignored. The rules of quantum
mechanics tell us that, to take the identity of particles into account,
we have to replace $M^{(0)}_N(q)$ by a new amplitude $M_N(q)$ which is a
sum over all permutations of the momenta $(q_1,...q_N)$ 
\begin{equation}
M^{(0)}_N(q) \rightarrow M_N(q) \equiv \sum_P M^{(0)}_N(q_P). \label{1}
\end{equation} 
This would be the end of the story if particle production
was described by a single matrix element. In general, however, we have
to average over parameters which are not measured and therefore the
correct description of the multiparticle final state is achieved in
terms of the density matrix 
\begin{equation} \rho^{(0)}_N(q,q')
=\sum_{\omega} M^{(0)}_N(q,\omega) M^{(0)*}_N(q',\omega), \label{2}
\end{equation}
 rather than in terms of a single production amplitude.
The sum in (\ref{2}) runs over all quantum numbers $\omega$ which are
not measured in a given situation. $\rho^{(0)}(q,q')$ gives all
available information about the system in question. At this point it is
useful to note that, when tranformed into (mathematically equivalent)
Wigner representation 
\begin{equation} W_N(\bar{q},x) = \int d(\Delta q)
e^{ix\Delta q} \rho^{(0)}_N(\bar{q},\Delta q) \label{3} 
\end{equation}
($\bar{q}= (q+q')/2; \;\; \Delta q =q-q'$) it gives information about
the distribution of momenta and positions of the particles (see, e.g.,
\cite{bk} for a discussion of this point).

 Using (\ref{1})  and (\ref{2}) one
easily arrives at the formula for the {\it corrected}
 (i.e., with identity of particles taken into account)
density matrix $\rho_N(q,q')$
  and one finally obtains the observed multiparticle density
\begin{equation}
\Omega_N(q)= \frac1{N!} \sum_{P,P'} \rho^{(0)}(q_P,q_{P'})   \label{4}
\end{equation}
where the sum runs over all permulations $P$ and $P'$ of the momenta
$(q_1,...q_N)$. The factor $\frac1{N!}$ appears because the phase space for $N$
identical particles is $N!$ times smaller than the phase space for $N$
non-identical particles. The formula (\ref{4}) is in common
use\footnote{Using the symmetry properties  of the density matrix, the
double sum in (\ref{4}) can be reduced to a single sum. The factor
$\frac1{N!}$ is then absent.}  and is the basis of our further discussion.

\subsection {A theoretical laboratory: independent particle production}

The case of independent particle production is an attractive theoretical
laboratory which, although not expected to
 describe all details of the data,
reveals -nevertheless- some  generic  features of the problem.
This was first recognized by Pratt \cite{pr}.
In terms of the density matrix, the independent production means that
the density matrix factorizes into a product of single-particle density
matrices
\begin{equation}
\rho^{(0)}_N(q,q') =
\rho^{(0)}(q_1,q_1')\rho^{(0)}(q_2,q_2')....\rho^{(0)}(q_N,q_N')
\label{5}
\end{equation}
and that the multiplicity distribution is the Poisson one
\begin{equation}
P^{(0)}(N) = e^{-\nu}\frac{\nu^N}{N!}.         \label{6}
\end{equation}

 It turns out \cite{zi,zh} that in the case of a
Gaussian density matrix  the problem can be solved analytically.
             The main results (valid
also in  the general case of an 
arbitrary density matrix \cite{bz}) can be
listed as follows.

(a) All correlation functions $K_p(q_1,...,q_p)$ and the single particle
 distribution $\Omega(q)$ can be
expressed in terms of one (hermitian)  function $L(q,q')=L^*(q',q)$
 of two momenta:
\begin{eqnarray}
\Omega(q)= L(q,q); \;\; K_2(q_1,q_2)=L(q_1,q_2)L(q_2,q_1);\nonumber \\
K_3(q_1,q_2,q_3)= L(q_1,q_2) L(q_2,q_3)L(q_3,q_1)+ L(q_1,q_3) L(q_3,q_2)
L(q_2,q_1), \label{7}
\end{eqnarray}
and analogous formulae for higher correlation functions.

(b) At very large phase-space density of particles, the distribution approaches
a singular point representing the phenomenon of Bose-Einstein
condensation: almost all particles populate the eigenstate of
$\rho^{(0)}(q,q')$ corresponding to the {\it largest eigenvalue}. The
resulting multiplicity distribution is very broad (almost flat) so that,
e.g., the probability of an event with no single $\pi^0$ produced is
non-negligible 
\footnote{This effect was also considered in connection
of the possible production of the Disoriented Chiral Condensate
\cite{blk,bj}. The present argument adds another obstacle on the
difficult road to observation of DCC. }.

It should be not surprizing if the very restrictive condition of
independent production, as expressed by (\ref{5},\ref{6}) is not
realized in nature. Nevertheless the comparison of the relations
(\ref{7}) with data is interesting, since they are a sort of reference
point allowing to judge if the observed multiparticle correlations are
"large" or "small" with respect to the observed two-body correlations.
The existing evidence leads to rather interesting, although
controversial, conclusions. At the Matrahaza meeting, Lorstad \cite{lo} 
demonstrated that there are practically no
genuine three-particle correlations\footnote{The importance of the
absence of 3-particle correlations in heavy ion collisions was
emphasized already some time ago \cite{is}.} in S-Pb collisions at CERN
SPS. Since the two-particle correlations are clearly visible, this
observation is not easy to  reconcile with Eq.({7}). It was 
earlier shown by
Eggers et al \cite{elb} that the UA1 data are also in contradiction with
(\ref{7}), although in this case the 3-body correlations seem to be too
large to satisfy (\ref{7}). On the other hand,
it was shown recently by Arbex et al \cite{arb}
  that
the NA22 data agree well with (\ref{7}). This striking difference between the
behaviour of heavy ion and "elementary" collisions is certainly very
interesting and deserves further attention.

We cannot thus consider the results obtained from (\ref{5},\ref{6}) to
be a realistic description of the data. Nevertheless, the main conclusion
about the possibility of Bose-Einstein condensation remains an
interesting option which is worth a serious consideration \cite{kz4,shu}.

\subsection { Monte Carlo simulations}

In this situation, the practical method to study the effects of BE
symmetrization on particle spectra is to implement it into the Monte
Carlo codes. A "minimal" method of performing this task was suggested
some time ago  \cite{bk}. The idea is to take an existing code (which
reproduces the distribution of particle momenta, i.e. the diagonal
elements of the density matrix) and to introduce an {\it ansatz} for the off-diagonal
elements of the multiparticle density matrix (\ref{2})\footnote{As seen
from (\ref{3}) this corresponds to introducing an - a priori arbitrary -
distribution of particle emision points in configuration space.}. Each
event generated by the MC code is then given a weight which is
calculated as the ratio of symmetrized distribution [Eq.(\ref{4})], and
the unsymmetrized one. In this way the modification of the original
spectra is kept at the minimum.

A practical realization of this idea (in its simplest version) has been
developped by the Cracow group \cite{fw} and shall be  presented by Fialkowski
at this meeting. They propose the unsymmetrized density matrix in the
form 
\begin{equation} \rho^{(0)}_N(q,q')= P_N(\bar{q}) \prod_{i=1}^N
w(q_i-q_i') \label{9} 
\end{equation}
 where $P_N(q)$ is the probability
of a given configuration obtained in JETSET and $w$ is a Gaussian. This
prescription does not modify the diagonal elements of the unsymmetrized
density matrix ($w(0)=1$) and, moreover, does not introduce any new
correlations between emission points of the produced particles (when
transformed into Wigner representation, Eq.(\ref{4}), the product $\prod w(q_i-q_i')$
becomes the product $\prod w(x_i)$). Thus (\ref{9}) can indeed be
considered as a minimal modification of the existing code. The autors
find that this prescription represents well the existing data on
two-particle correlations and that they can recover the experimental
multiplicity distribution by a simple rescaling with the formula $P(N)
\rightarrow P(N)cV^N$, without the necessity of refitting the JETSET
parameters.

They also  studied  production of pair of $W$ bosons at LEP
II \cite{fw} and found fairly strong effects of quantum interferrence.
One may note, however, that 
  in present version of the
model the position of particle emission point is not correlated with its
momentum, whereas such correlation  is likely to be present in reality.  Consequently, the obtained results 
may be overestimating the effect. This
deficciency is easy to repair \cite{bk} but on the prize of introducing more
parameters.

A more fundamental approach has been pursued since some time by
Andersson and Ringner \cite{ar} and shall be presented here
by Todorova-Nova. It is based on the  paper by
Andersson and Hoffman \cite{ah}.
 In this case the "uncorrected" matrix element
represents the decay of one Lund string
 which is then symmetrized according to the procedure explained
in introduction. Two particle correlations are well described and
several interesting effects are predicted. Among them: (a) the
longitudinal and transverse correlations are expected to be different
because they are controlled by two different physical mechanisms; (b)
Three particle correlations are predicted non-vanishing and were
actually calculated; (c) $WW$ production was studied and no significant
mass shift is expected; (d) No multiplicity shift in the $W$ decay is
predicted.

This last conclusion is a consequence of the fact that, in case of more
than one string present in the final state, no symmetrization between
particles stemming from different strings is performed. This corresponds
to the assumption that the strings are created at a very large distance
from each other. One thus may expect that in a more realistic treatment
some multiplicity shift should be present
\footnote{A  contribution to this problem was presented recently by B.Buschbeck et al. \cite{bbu,abm}.}

Finally, let me add that in both \cite{fw} and
\cite{ar}  the "interconnection effect" \cite{el} (which has tendency to reduce the
multiplicity) is neglected. The full phenomenological analysis of the
data is therefore certainly more complicated, as we shall hear   from de Jong.

\subsection { Probing the space-time structure}

Much attention is also devoted to the information one may
obtain from the data on quantum interference about the space-time
structure of the multiparticle system created in the collision 
[c.f.Eq.(\ref{4})]. Although
such analyses  have  a somewhat  limited scope, as they (i) provide only
 information about the system at the freeze-out and (ii)
require several additional assumptions - they give
nevertheless a unique opportunity to investigate this problem. Most of
the {\it caveats} are thus usually postponed to the future (and better
data) and the analysis is carried on.

The recently presented investigations were based on the hydrodynamic
approach. Some of them  were discussed here already
during the Session on Heavy Ion Interactions. The main features 
\footnote{A more detailed description is given in \cite{abm}.} are: (i)
The shape of the particle emission region is consistent with the in-out
scenario of Bjorken \cite{bj2};
 (ii) The longitudinal size of the "fireball" from
which a bulk of particles are emitted is several times larger in heavy
ion collisions than in hadron-hadron interactions; (iii) Particle
emission process starts rather late in heavy ion collisions (after about
4 fm in S-Pb interactions \cite{}), as compared to the elementary
collision where it happens immediately after collision \cite{}; (iv) the
emission process, once started,  does not last for a long time:
less than 2 fm for elementary and about 3 fm for S-Pb interactions.

These features clearly indicate that a heavy ion collision is
 indeed followed by 
creation of an longitudinally expansing "fireball" in which
 some kind of matter is "boiling" for a considerable time. Once it is
sufficiently cooled, however, its decay is rather fast. It is presumably
too early to claim that it is formed from the quark-gluon plasma but
nevertheless this behaviour is rather suggestive.

\section {Intermittency}

Intermittency \cite{bp}, postulates scaling of the
multiparticle spectra. A rather complete review of the subject 
is now available \cite{ddk}, therefore I shall restrict myself 
to few remarks expressing my personal view on the progress
achieved in the last decade.

 (i) The scaling hypothesis can be formulated in many ways
and, indeed, much work was devoted to improvements and generalizations
of the original proposal \cite{bp} expressed in terms of the (normalized)
factorial moments 
\begin{equation}
F_q \equiv <n(n-1)...(n-q+1)>/<n>^q \sim (bin\;\; size)^{-f_q}.    \label{12}
\end{equation}
The result was an impressive progress in the developpment of more
 sophisticated tools which 
are much better suited for investigation of many, sometimes very
detailed, aspects of the problem. Let me particularly emphasize 
the importance of the correlation integrals,
first introduced in \cite{lipa}.
Their use was decisive in improving accuracy of the data and thus
to substantiate the evidence for the effect.

 (ii) On the experimental side, the analysis of high precision data 
(particularly those of NA22 and UA1 experiments) allowed to establish
-beyond a reasonable doubt- a close connection between intermittency and
Bose-Einstein correlations, as suggested \cite{car,gyu} 
almost immediately after
first experimental evidence  for increasing factorial
moments. This observation allowed to
understand the scaling of the momentum spectra as a reflection of the
-more fundamental- scaling in configuration space \cite{abg}. 
I have impression that the importance of this fact is not yet fully
appreciated.

(iii) Recently, a general solution of the model of multiplicative
cascade was obtained \cite{gre}. One can thus hope for a significant
progress in understanding the scaling phenomenon.

 (iv) Finally, let me also mention  another interesting development, namely 
the generalization of the notion of scaling by  the idea of 
self-affinity \cite{wuy}. I think it is an interesting direction to pursue
and I hope that more data on the subject shall be soon available.We 
shall hear more about this from Liu.

The major disappointment I see after all these years is that -in fact-
no convincing theoretical basis was found
 for the phenomenon of intermittency, although  it  seems to be indeed 
an universal  feature of particle spectra \cite{ddk,fiu}
\footnote{The second order phase transition was invoked by many authors \cite{abh}
as a possible explanation. This  is certainly a valid idea 
but it does not explain 
universality of the phenomenon.}.
Does it mean
that the simple effect one observes 
is a purely accidental result of summation of much more  complex 
contributions? Perhaps. Nevertheless, I am convinced that
 the search for a more fundamental
reason of the apparent scaling in particle spectra is worth to continue.

\section {QCD and multiparticle correlations}

 It is now rather well established
that the average multiplicity and single particle spectra are well
described by perturbative QCD supplemented with the principle of
parton-hadron duality \cite{phd}.

 In my opinion, at present, the real challenge to the idea of parton-hadron duality 
 is to explain
the data on differential correlation functions \cite{abg}. Indeed, it is hard to
understand how the momenta of the produced hadrons can follow so closely
the momenta of the created partons that the correlations between them
are not washed out\footnote{This problem is much less serious if
one considers only the integrated correlation functions. In this connection,
see the discussion at the recent meeting on Correlations
 and Fluctuations, Matrahaza, June 1998 and contibutions to this session
 by Metzger and Chekanov.}   .
 Therefore a
non-trivial extension of the principle of parton-hadron duality must be
formulated in order to give quantitative meaning to perturbative
calculations of multiple production. It was therefore 
rather recomforting to learn  
that indeed the predictions of
perturbative QCD formulated some time ago \cite{pqcd}, are badly
violated by the L3 data \cite{ki}. On the other hand, the same data are well
described by the JETSET code. The conclusion is that the hadronization
part is  not correctly taken into account by the simple (naive?)
parton-hadron duality. More about that later in this Session by Mandl.

This is not to say that the subject is closed: Ochs \cite{oc} pointed
out that the tested QCD calculations 
included several simplifying assumptions (the most important among them
seems the neglect of energy-momentum consevation) and thus it is not
obvious  which part of the result is actually responsible for the
failure.    It is clear, nevertheless, that further work on these lines
must seriously address the problem of parton-hadron duality and its
range of application.

\section {Event-by-event analysis}

Event-by-event analysis clearly emerges as a next logical step in
studies of multiparticle fluctuations. The subject is not yet well
developped, however, and neither the physics nor methods  sufficiently 
understood to define 
precisely what we are really searching for. Therefore, I can only list
a few ideas of potential interest. I am fully aware that some of them
may not work and that others, more interesting, may well 
  be proposed in the near future.

There are two basic reasons why event-by event fluctuations attract
attention. The first one, more spectacular, is to look for large
deviations of some events from the average, with the hope of finding 
a hitherto unobserved effect. The second,
 more pragmatic, is to  measure the distribution of a quantity
defined for a single event and thus obtain  
additional information,  helping to understand the physics of
the process. This is well illustrated by multiplicity distribution
which is the simplest event-by-event analysis one may think of. It was 
studied since long time \footnote{ Contributions related to multiplicity distributions
shall be presented by Hegyi, Ploszajczak and Blazek.} and was of great help in understanding the 
physics of multiparticle production.

Let me now go to my list:

(i) Recently, Stodolsky \cite{sto} proposed to study fluctuations in
transverse momentum (see also \cite{shu}). The idea is that, if the transverse momentum
distribution in an event can be related to its "temperature", one 
obtains in this way the distribution of "temperatures" of the events.
Now, if 
thermodynamics is a correct decription of the process in question, the fluctuations of temperature
can be related to to heat capacity $C_V$ of the system \cite{lan}:
\begin{equation}
\frac{(\Delta T)^2}{T^2} = \frac{k}{C_V}   \label{13}
\end{equation}
where $k$ is the Boltzmann constant. This obviously may be very helpful 
in searching for phase transition. Even far from phase transition, however, a measurement
of this kind can provide a lot of information on (a) the properties 
of the system in question and (b) 
whether it is indeed close to thermodynamic equlibrium. To take a simple example:
In case of ideal gas one has $ E= C_V T$, where $E$ is the energy of the system.
We thus obtain
\begin{equation}
\frac{(\Delta T)^2}{T^2} = \frac{kT}{E}.   \label{a13}
\end{equation}
The point is that both L.H.S. and R.H.S. of this equation can be measured and thus one 
may hope to estimate the deviation of the system from the ideal gas approximation.
\footnote{Recently, the first results on temperature fluctuations
were presented by NA49 coll. \cite{rol}.}

(ii) Another important issue was raised by Hwa \cite{hw2}. He pointed out the
essential difference between the determination of  fractal parameters
in case of dynamical systems and in case of systems of many particles.
In the dynamical system one can generate the {\it time sequence} and
thus estimate how fast the different trajectories diverge. In case of
multiparticle systems we do not have a time sequence and thus we have to
rely on {\it patterns}. The question in this case is: how different are
the patterns of different events. Hwa proposed to {\it measure} the
pattern of an event by the factorial moment associated with it. One can
then ask the question how this measure fluctuates from event to event.
Studying moments of this distribution provides a measure of
event-to-event fluctuation\footnote{To study the moments of the
factorial moments was suggested already some time ago \cite{bzd}.}. When
they are considered as function of bin size, it is possible to define
appropriate fractal dimensions which conveniently summarize the
information. For the details the reader is referred to the original
paper \cite{hw2}. I personally feel that this is an important conceptual
step in our thinking about the problem, although I am not fully
convinced that the proposed measure cannot be improved.

(iii)  The studies of possible phase transitions in the multiparticle systems 
produced in high-energy collisions \cite{anto}
suggest that the fractal behaviour may strongly
fluctuate from one event to another. It follows that it is       
essential to be able to study the  fractal behaviour in
event-by-event analysis. The feasibility of this program was
investigated recently \cite{zi2}. It seems to be rather promising.

(iv) The factorial moments can also be considered as a very sensitive
signature for clustering of particles in small bins of momentum
phase-space.
Indeed, a factorial moment of order $q$ is sensitive only to the
clusters containig at least $q$ particles. This obviously eliminates
very effectively any background. This point was recently illustrated
by KLM collaboration analysing the collisions of 160 GeV/A Pb nucleons 
in emulsion \cite{KLM}. They found that events may differ drastically in the
behaviour of their factorial moments, while no great difference is seen
when they are analysed by other methods.

(v) 
 The distribution of the HBT
radii obtained from individual events was also recognized 
since some time as a very
 interesting object to study. Recently, first data on this subject
were presented  by NA49 collaboration \cite{sey,abm}.
 Although statistics is still limited (and the authors
themselves do not attach too much meaning to the details of the plot)
the results  clearly show that the measurement is feasible and we may well hope
for some exciting news in not-too-distant future.

\section* {Acknowledgements}

I would like to thank N.Antoniou and C. Ktorides
 for the kind hospitality at Delphi. The encouragement from W.Kittel
is highly appreciated.
This work was supported in part by the KBN Grant No 2 P03B 086 14.


\end{document}